\documentclass[aps,pra,superscriptaddress,showpacs]{revtex4}%
\usepackage{amsfonts}
\usepackage{amsmath}
\usepackage{amssymb}
\usepackage{graphicx}%
\begin{document}
\title{Beyond Fermi pseudopotential: a modified GP equation}
\author{Haixiang Fu}
\affiliation{Shanghai Institute of Optics and Fine Mechanics, Chinese Academy of Sciences,
Shanghai 201800, China}
\author{Yuzhu Wang}
\affiliation{Shanghai Institute of Optics and Fine Mechanics, Chinese Academy of Sciences,
Shanghai 201800, China}
\author{Bo Gao}
\affiliation{Shanghai Institute of Optics and Fine Mechanics, Chinese Academy of Sciences,
Shanghai 201800, China}
\affiliation{Department of Physics and Astronomy, University of Toledo, Toledo, Ohio 43606}
%\thanks{Corresponding author}

\date{February 17, 2003}

\begin{abstract}
We present an effective potential and the corresponding modified
Gross-Pitaevskii equation that account for the energy dependence
of the two-body scattering amplitude through an effective-range expansion.
For the ground state energy of a trapped condensate, the theory leads 
to what we call a shape-dependent confinement correction that improves 
agreements with diffusion Monte Carlo calculations. 
The theory illustrates, for relatively strong confinement and/or high density,
how the shape dependence on atom-atom interaction 
can come into play in a many-atom quantum system.
  
\end{abstract}


\maketitle

\section{Introduction}

The Gross-Pitaevskii equation (GPE) has in recent years played a central role
in our understanding of gaseous Bose-Einstein condensates (BEC) \cite{dal99,leg01}.
Its astonishing success has to do with the fact that in
most experiments to date, both the gaseous parameter,
$na^3$ (here $n$ is the number density and $a$ is the scattering length), 
and the parameter $a/L_0$, 
which characterizes the strength of confinement 
(with $L_0$ being a measure of the size of the trap), have been very small. 
Recently, a new breed of experiments \cite{feshbe} have
emerged that take advantage of Feshbach resonances in cold atomic
collision \cite{tie92} to make the scattering length $a$ tunable 
through a magnetic field. 
Such experiments promise to take us into parameter regimes 
in which the validity of GPE and its underlying assumptions, such as the 
shape-independent pseudopotential approximation \cite{dal99,leg01}, 
have to be carefully examined.

Most of existing theories going beyond the GPE \cite{bra97,fab99,blu01}
have focused on quantum fluctuation that is responsible for
the lowest order correction in $na^3$.
But it has become clear that eventually, 
%for sufficiently large $na^3$ and/or sufficiently strong confinement,
a better description of atom-atom interaction, beyond the 
pseudopotential approximation that is the basis of
the GP theory, will be required.
For homogeneous systems, this has been demonstrated in
a recent work by Cowell \textit{et al.} \cite{cow02}, 
which shows that different potentials
having the same scattering length $a$ can lead to vastly different
ground state energy for $na^3$ of the order of 0.05 or greater.
For inhomogeneous systems, recent works on two atoms in a trap
have shown that the shape-independent approximation becomes less valid 
under strong confinement \cite{tie00,bol02,blu02}.

Neither of these results are surprising. The shape-independent approximation
assumes the two-body scattering amplitude to be a constant over
the momentum (or energy) range of interest. 
For larger $na^3$ or stronger confinement, a greater range of momentum
states become involved and this assumption becomes less applicable.
This point has already been illustrated for two-atom systems under
strong confinement, where it has been shown
that inclusion of the energy-dependence of scattering amplitude 
leads to much improved results for the energy levels of the
system \cite{tie00,bol02,blu02}. The goal of this article is to show 
how a similar idea can be generalized to a N-body system.

In light of the success of the GP theory \cite{dal99,leg01}, we focus here on
a perturbative approach around the GPE. An effective potential is proposed that
incorporates the energy dependence of the two-body scattering amplitude through
an effective-range expansion. A corresponding modified GP equation is derived.
By taking advantage of the relationship between effective range 
and scattering length as implied by the quantum-defect 
theory (QDT) \cite{gao98b,gao01}, no new inputs 
from scattering calculations are required other than the standard 
scattering length and the $C_6$ coefficient.
For the ground state energy of an inhomogeneous system, our numerical results
show that the modified equation leads to better agreements with diffusion 
Monte Carlo (DMC) results \cite{blu01} than theories that consider only the quantum 
fluctuation correction. More importantly, the theory introduces the concept
of shape-dependent confinement correction, an origin of shape-dependence 
on the interaction potential that is absent in a homogeneous system.

\section{Modified GP Equation}

In a second-quantization formulation, a system of N bosons in an external
trapping potential $V_{\text{ext}}(\mathbf{r})$ is described by a Hamiltonian
\begin{widetext}
\begin{equation}
\widehat{H} = \int d\mathbf{r}\,\widehat{\Psi}^{\dagger}(\mathbf{r}%
)\left[  -\frac{\hbar^{2}}{2m}\nabla^{2}+V_\text{ext}(\mathbf{r})\right]  \widehat{\Psi
}(\mathbf{r}) + \frac{1}{2}\int d\mathbf{r}_{1}\int d\mathbf{r}_{2}\,\widehat{\Psi
}^{\dagger}(\mathbf{r}_{1})\,\widehat{\Psi}^{\dagger}(\mathbf{r}_{2}%
)V(\mathbf{r}_{1}-\mathbf{r}_{2})\widehat{\Psi}(\mathbf{r}%
_{2}\mathbf{)\,}\widehat{\Psi}(\mathbf{r}_{1}) \;.
\end{equation}
\end{widetext}
Here $V$ is the interaction between particles. $\widehat{\Psi
}(\mathbf{r})$ and $\widehat{\Psi}^{\dagger}(\mathbf{r})$ are the bosonic
annihilation and creation operators that satisfy the commutation relations
\begin{align}
\lbrack\widehat{\Psi}^{\dagger}(\mathbf{r}_{1}),\widehat{\Psi}^{\dagger
}(\mathbf{r}_{2})]  &  =[\widehat{\Psi}(\mathbf{r}_{1}),\widehat{\Psi
}(\mathbf{r}_{2})]=0,\\
\lbrack\widehat{\Psi}(\mathbf{r}_{1}),\widehat{\Psi}^{\dagger}(\mathbf{r}%
_{2})]  &  =\delta(\mathbf{r}_{1}-\mathbf{r}_{2}).
\end{align}

Taking the inter-particle potential $V$ to be that of a mean field
\begin{equation}
V_{\text{MF}}\ =g\delta(\mathbf{r}_{1}-\mathbf{r}_{2}), \label{eq:mfpot}%
\end{equation}
where
\begin{equation}
g=\frac{4\pi\hbar^{2}a}{m}\;,
\end{equation}
and ignoring quantum fluctuation, one arrives at the usual GP equation
\cite{dal99,leg01}
\begin{equation}
i\hbar\frac{\partial}{\partial\,t}\,\Phi(\mathbf{r},t\mathbf{)}=\left[
-\frac{\hbar^{2}}{2m}\nabla^{2}+V_{\text{ext}}(\mathbf{r})+g\left\vert
\Phi\right\vert ^{2}\right]  \Phi(\mathbf{r},t)\;.
\end{equation}
Here $\Phi=\langle\widehat{\Psi}\rangle$ and is nomalized by $\int d\mathbf{r}%
|\Phi|^{2}=N$, with $N$ being the total number of particles.

From a two-body scattering point of view, the mean-field potential, given by
Eq.~(\ref{eq:mfpot}), is such that it gives, in the first Born approximation,
the correct foward scattering amplitude at zero energy. It is a
shape-independent approximation that ignores completely the energy-dependence
of the scattering amplitude. While this may be a good approximation for small
values of $na^{3}$ and for weak confinement, it becomes less valid 
for strong confinement and/or greater $na^{3}$. This is true even at zero
temperature, as confinement and/or quantum depletion invariably involve
nonzero momentum states.

For a better treatment of atomic interaction, while preserving much of the
structure of the GP theory, an effective interaction is proposed here that
incorporates the energy dependence of the scattering amplitude through an
effective-range expansion. Specifically, we look for a $\widehat{V}_{\text{eff}}$
that gives, in the first Born approximation, the correct real part of
the \textit{forward} scattering amplitude,
\begin{equation}
-\frac{m}{4\pi\hbar^{2}}\int d\mathbf{r}\:e^{-i\mathbf{k}\cdot\mathbf{r}%
}\widehat{V}_{\text{eff}}\:e^{+i\mathbf{k}\cdot\mathbf{r}}=\text{Re}%
[f(k,\theta=0)], \label{eq:cveff}%
\end{equation}
to the order of $k^{2}$ (the first order in energy). This requirement 
basically ensures a better optical potential for an atom moving in the
medium of others \cite{rod67}.

For the range of energies of interest here, only the $s$ wave scattering is
important. In this case, the scattering amplitude can be written in terms of
the $s$ wave phase shift $\delta_{0}(k)$ as
\begin{equation}
f(k,\theta)=\frac{1}{k}\,\frac{1}{\cot\delta_{0}(k)-i}. \label{eq1}%
\end{equation}
From the standard effective-range expansion
\begin{equation}
k\cot\delta_{0}(k)=-\frac{1}{a}+\frac{1}{2}\,r_{e}\,k^{2}+..., \label{eq2}%
\end{equation}
where $r_{e}$ is the effective range, we have, to the order of $k^{2}$,
\begin{equation}
\text{Re}[f(k,\theta)]=-a+a^{2}(a-\frac{1}{2}r_{e})k^{2}. \label{eq3}%
\end{equation}

It is straightforward to verify that the $\widehat{V}_{\text{eff}}$ satisfying
Eq.~(\ref{eq:cveff}) can be written as
\begin{equation}
\widehat{V}_{\text{eff}}\ =V_{\text{MF}}(\mathbf{r}_{1}-\mathbf{r}%
_{2})+\widehat{V}_{\text{mod}}\text{\thinspace}(\mathbf{r}_{1}-\mathbf{r}%
_{2}),
\end{equation}
where
\begin{equation}
\widehat{V}_{\text{mod}}\ =\frac{g_{2}}{2}\left[  \delta(\mathbf{r}%
_{1}-\mathbf{r}_{2})\nabla_{\mathbf{r}_{1}-\mathbf{r}_{2}}^{2}+\nabla
_{\mathbf{r}_{1}-\mathbf{r}_{2}}^{2}\delta(\mathbf{r}_{1}-\mathbf{r}%
_{2})\right]  \;, \label{eq5}%
\end{equation}
and
\begin{equation}
g_{2}=\frac{4\pi\hbar^{2}a^{2}(a-\frac{1}{2}r_{e})}{m}\;.
\end{equation}
This potential is consistent with the pseudopotential expansion of
Huang and Yang \cite{hua57}, except that it is Hermitian and is 
represented in the coordinate space.

With this effective interaction, the many-body Hamiltonian becomes
%TCIMACRO{\TeXButton{beginwide}{\begin{widetext}}}%
%BeginExpansion
\begin{widetext}%
%EndExpansion%
\begin{align}
\widehat{H}  &  \equiv\widehat{H}_{\text{MF}}+\widehat{H}_{\text{mod}%
}\nonumber\\
&  \equiv\left\{  \int d\mathbf{r}\,\widehat{\Psi}^{\dagger}(\mathbf{r}%
)\left[  -\frac{\hbar^{2}}{2m}\nabla^{2}+V_\text{ext}(\mathbf{r})\right]  \widehat{\Psi
}(\mathbf{r)+}\frac{1}{2}\int d\mathbf{r}_{1}\int d\mathbf{r}_{2}%
\,\widehat{\Psi}^{\dagger}(\mathbf{r}_{1})\,\widehat{\Psi}^{\dagger
}(\mathbf{r}_{2})V_{\text{MF}}\,(\mathbf{r}_{1}-\mathbf{r}_{2})\widehat{\Psi
}(\mathbf{r}_{2}\mathbf{)\,}\widehat{\Psi}(\mathbf{r}_{1}\mathbf{)}\right\}
\nonumber\\
&  +\frac{1}{2}\int d\mathbf{r}_{1}\int d\mathbf{r}_{2}\,\widehat{\Psi
}^{\dagger}(\mathbf{r}_{1})\,\widehat{\Psi}^{\dagger}(\mathbf{r}_{2}%
)\widehat{V}_{\text{mod}}\,(\mathbf{r}_{1}-\mathbf{r}_{2})\widehat{\Psi
}(\mathbf{r}_{2}\mathbf{)\,}\widehat{\Psi}(\mathbf{r}_{1}\mathbf{)}.
\label{eq6}%
\end{align}%
%TCIMACRO{\TeXButton{endwide}{\end{widetext}}}%
%BeginExpansion
%\end{widetext}%
%EndExpansion
The term associated with $\widehat{V}_{\text{mod}}$ can be simplified in the
two-body centre-of-mass frame [$\mathbf{r}=\mathbf{r}_{1}-\mathbf{r}_{2}$,
$\mathbf{R}=(\mathbf{r}_{1}+\mathbf{r}_{2})/2$], in which $\widehat{H}_{\text{mod}}$ 
takes the form:%
%TCIMACRO{\TeXButton{beginwide}{\begin{widetext}}}%
%BeginExpansion
%\begin{widetext}%
%EndExpansion
%
\begin{equation}
\widehat{H}_\text{mod} = \frac{1}{4}g_2\int d\mathbf{R}\int d\mathbf{r}\,\widehat{\Psi}%
^{\dagger}(\mathbf{R+}\frac{\mathbf{r}}{2})\,\widehat{\Psi}^{\dagger
}(\mathbf{R-}\frac{\mathbf{r}}{2})\left[ \delta(\mathbf{r}%
)\,\nabla_{\mathbf{r}}^{2}+\nabla_{\mathbf{r}}^{2}\,\delta(\mathbf{r})\right]  
\widehat{\Psi}(\mathbf{R-}\frac{\mathbf{r}}{2}\mathbf{)\,}%
\widehat{\Psi}(\mathbf{R+}\frac{\mathbf{r}}{2}\mathbf{).} \label{eq7}%
\end{equation}
%
%TCIMACRO{\TeXButton{endwide}{\end{widetext}}}%
%BeginExpansion
\end{widetext}%
%EndExpansion

After integrating over $\mathbf{r}$ and some integrations by parts,
$\widehat{H}_{\text{mod}}$ can be written as
\begin{equation}
\widehat{H}_{\text{mod}}=\frac{1}{4}g_{2}\int d\mathbf{R}\,\widehat{\Psi
}^{\dagger}(\mathbf{R})\left[  \nabla^{2}\left(  \widehat{\Psi}^{\dagger
}(\mathbf{R})\widehat{\Psi}(\mathbf{R)}\right)  \right]  \widehat{\Psi
}(\mathbf{R)}. \label{eq8}%
\end{equation}
Note that the Laplacian operator operates only on the number-density
operator $\widehat{\Psi}^{\dagger}(\mathbf{R})\widehat{\Psi}(\mathbf{R)}$.
Simplification of $\widehat{H}_{\text{MF}}$ follows the usual procedure, and
leads finally to
\begin{align}
\widehat{H}  &  =\int d\mathbf{R}\,\widehat{\Psi}^{\dagger}(\mathbf{R})\left[
-\frac{\hbar^{2}}{2m}\nabla^{2}+V_{\text{ext}}(\mathbf{R})\right]
\,\widehat{\Psi}(\mathbf{R})\nonumber\\
&  +\frac{1}{2}\int d\mathbf{R}\,\widehat{\Psi}^{\dagger}(\mathbf{R})\,\left[
g\left(  \widehat{\Psi}^{\dagger}(\mathbf{R})\widehat{\Psi}(\mathbf{R}%
)\right)  \right. \nonumber\\
&  \left.  +\frac{1}{2}g_{2}\nabla^{2}\left(  \widehat{\Psi}^{\dagger
}(\mathbf{R})\widehat{\Psi}(\mathbf{R})\right)  \right]  \widehat{\Psi
}(\mathbf{R}). \label{eq:Ham}%
\end{align}
From this Hamiltonian, both the static and the dynamic properties of a
many-atom Bose system can be studied. We focus here on the ground 
state energy of the system.

Ignoring both quantum fluctuation and $\widehat{H}_{\text{mod}}$,
Eq.~(\ref{eq:Ham}) leads to the standard GP energy functional
\begin{equation}
E_{\text{GP}}\left[  \Phi\right]  =\int d\mathbf{r}\left[  \frac{\hbar^{2}%
}{2m}\left\vert \nabla\Phi\right\vert ^{2}+V_{\text{ext}}(\mathbf{r}%
)\left\vert \Phi\right\vert ^{2}+\frac{1}{2}g\left\vert \Phi\right\vert
^{4}\right]  , \label{eq11}%
\end{equation}
which corresponds to a time-independent GP equation for the ground-state
wave function with a chemical potential $\mu$%
\begin{equation}
\left[  -\frac{\hbar^{2}}{2m}\nabla^{2}+V_{\text{ext}}(\mathbf{r})+g\left\vert
\Phi\right\vert ^{2}\right]  \Phi(\mathbf{r},t)=\mu\Phi(\mathbf{r},t).
\label{eq13}%
\end{equation}

Inclusion of the lowest-order term due to quantum 
fluctuation leads to an energy functional \cite{lhy57,fab99}
\begin{align}
E_{\text{MGPI}}\left[  \Phi\right]   &  =\int d\mathbf{r}\left[  \frac
{\hbar^{2}}{2m}\left\vert \nabla\Phi\right\vert ^{2}+V_{\text{ext}}%
(\mathbf{r})\left\vert \Phi\right\vert ^{2}\right. \nonumber\\
&  \left.  +\frac{1}{2}g\left\vert \Phi\right\vert ^{4}\left(  1+\frac
{128}{15}\frac{1}{\sqrt{\pi}}a^{3/2}\left\vert \Phi\right\vert \right)
\right]  ,
\end{align}
where the extra term due to quantum fluctuation was first derived by 
Lee, Huang and Yang \cite{lhy57}, and will be called the LHY term.
This functional corresponds to a modified GP equation that we call 
MGPI \cite{fab99,blu01}
\begin{equation}
\left[  -\frac{\hbar^{2}}{2m}\nabla^{2}+V_{\text{ext}}(\mathbf{r})+g\left\vert
\Phi\right\vert ^{2}+g_{1}\left\vert \Phi\right\vert ^{3}\right]
\Phi(\mathbf{r},t)=\mu\Phi(\mathbf{r},t), \label{eq:MGPI}%
\end{equation}
where
\begin{equation}
g_{1}=g\frac{32}{3\sqrt{\pi}}a^{3/2}.
\end{equation}

Inclusion of both quantum fluctuation and $\widehat{H}_{\text{mod}}$ to the
lowest order leads to an energy functional
\begin{align}
E_{\text{MGPII}}\left[  \Phi\right]   &  =\int d\mathbf{r}\left[  \frac
{\hbar^{2}}{2m}\left\vert \nabla\Phi\right\vert ^{2}+V_{\text{ext}}%
(\mathbf{r})\left\vert \Phi\right\vert ^{2}\right. \nonumber\\
&  \left.  +\frac{1}{2}g\left\vert \Phi\right\vert ^{4}+\frac{2}{5}%
g_{1}\left\vert \Phi\right\vert ^{5}+\frac{1}{4}g_{2}\left\vert \Phi
\right\vert ^{2}\nabla^{2}\left(  \left\vert \Phi\right\vert ^{2}\right)
\right]  .
\label{eq:MGPII1}
\end{align}
The corresponding modified GP equation, which we call MGPII, is
\begin{align}
&  \left[  -\frac{\hbar^{2}}{2m}\nabla^{2}+V_{\text{ext}}(\mathbf{r}%
)+g\left\vert \Phi\right\vert ^{2}+g_{1}\left\vert \Phi\right\vert ^{3}\right.
\nonumber\\
&  \left.  +\frac{g_{2}}{2}\nabla^{2}\left(  \left\vert \Phi\right\vert
^{2}\right)  \right]  \Phi(\mathbf{r},t)=\mu\Phi(\mathbf{r},t). 
\label{eq:MGPII2}%
\end{align}

MGPII incorporates the shape dependence on the interaction potential through
the effective range $r_{e}$. This additional parameter does not,
however, add much extra complexity to the theory, as $r_{e}$ and $a$ are generally
related. For a hard sphere potential, we have the well-know relation
\begin{equation}
r_{e}=(2/3)a. \label{eq:rehs}%
\end{equation}
For atoms with van der Waals interaction, $r_{e}$ can be determined from $a$
by \cite{gao98b}
\begin{equation}
r_{e}/\beta_{6} =\left(  \frac{2}{3x_{e}}\right)\frac{1}{(a/\beta_{6})^{2}}\left\{
1+\left[  1-x_{e}(a/\beta_{6})\right]^{2}\right\} ,
\label{eq:reinvr6}%
\end{equation}
where $\beta_{6}=\left(  mC_{6}/\hbar^{2}\right)  ^{1/4}$ is a length scale
associated with the van der Waals interaction, and $x_{e}\equiv\left[
\Gamma(1/4)\right]  ^{2}/\left(  2\pi\right)  $.
Similar relations for other long-range potentials should also exist,
an assertion that can be deduced from a more general QDT 
consideration \cite{gao01}.

The physical meaning of the correction due to
$\widehat{H}_{\text{mod}}$ can be better understood by estimating 
the order-of-magnitude of each term in Eq.~(\ref{eq:MGPII1}) or (\ref{eq:MGPII2}).
Consider a system of $N$ bosons confined to a length scale $L_0$.
The order-of-magnitude for the number density is $n\sim N/(L_0)^3$. 
Let $\epsilon_\text{MF} \sim na^3(\hbar^2/2m)(1/a)^2$ to represent the
energy per particle due to the mean field. It is easy to show that
the LHY quantum fluctuation term is of the order of 
$\epsilon_\text{LHY}\sim \epsilon_\text{MF}(na^3)^{1/2}$,
while the $\widehat{H}_{\text{mod}}$ term is the order of
$\epsilon_\text{mod}\sim \epsilon_\text{MF}(a/L_0)^{2}$.
It is thus clear that the two corrections are of different origin. 
The $\widehat{H}_{\text{mod}}$ correction has an order-of-magnitude that is
determined primarily by the strength of confinement. 
It depends also on the shape of the interaction 
potential through $g_2$ and $r_e$.
We will call this correction the shape-dependent confinement correction.
It is a new source of shape dependence that is absent in a homogeneous system 
($L_0\rightarrow \infty$ with $n$ being fixed). 
In this regard, it differs substantially from the shape dependence 
that can come at higher densities from two-body correlation at shorter 
length scales \cite{cow02,gao03}. 

A further comparison of the magnitudes of the LHY and 
the shape-dependent confinement corrections 
reveals other interesting features. From the estimates above, we have
$\epsilon_\text{mod}/\epsilon_\text{LHY} \sim (a/L_0)^{2}/(na^3)^{1/2}
\sim (a/L_0)^{1/2}/N^{1/2}$.
This means, for instance, that relative to the LHY correction,
the shape-dependent confinement correction is more important for
small $N$ than for large $N$, a point that will be further
demonstrated numerically.

We have chosen, up to this point, a formulation that most
closely parallels the standard GP formulation \cite{dal99,leg01}. 
For a small particle number
$N$, a number-conserving Hartree-Fock formulation is in fact more 
appropriate \cite{esr97,leg01}.
The corresponding results are easily derived.
For example, at the level of MGPII, we have
\begin{align}
E_{\text{MGPII}}\left[  \phi\right]   & = N\int d\mathbf{r}\left[  \frac
{\hbar^{2}}{2m}\left\vert \nabla\phi\right\vert ^{2}+V_{\text{ext}}%
(\mathbf{r})\left\vert \phi\right\vert ^{2}\right. \nonumber\\
&  \left.  +\frac{1}{2}g(N-1)\left\vert \phi\right\vert ^{4}+\frac{2}{5}%
g_{1}(N-1)^{3/2}\left\vert \phi\right\vert ^{5} \right. \nonumber\\
& \left. +\frac{1}{4}g_{2}(N-1)\left\vert \phi
\right\vert ^{2}\nabla^{2}\left(  \left\vert \phi\right\vert ^{2}\right)
\right]  .
\label{eq:MGPIIhf1}%
\end{align}
and
\begin{align}
&  \left[ -\frac{\hbar^{2}}{2m}\nabla^{2}+V_{\text{ext}}(\mathbf{r})
	+g(N-1)\left\vert \phi\right\vert ^{2}
	+g_{1}(N-1)^{3/2}\left\vert \phi\right\vert ^{3}\right.\nonumber\\
&  \left.  +\frac{g_{2}}{2}(N-1)\nabla^{2}
	\left(\left\vert \phi\right\vert^{2}\right)\right]
	\phi(\mathbf{r},t)=\mu\phi(\mathbf{r},t) ,
\label{eq:MGPIIhf2}%
\end{align}
where $\phi$ is normalized according to $\int d\mathbf{r}|\phi|^{2}=1$.
It is the Hartree-Fock formulation that we actually implement in our
numerical calculations, but the difference is small for large $N$ 
(between $N$ and $N-1$). Setting $g_2$, or both $g_2$ and $g_1$, to zero
in Eqs.~(\ref{eq:MGPIIhf1}) and (\ref{eq:MGPIIhf2}) leads to Hartree-Fock
results in MGPI and GP, respectively.

\section{Numerical Results}

We present here our numerical results for the ground state energy
of a BEC in different approximations. They are obtained using
an imaginary time evolution method with ADI algorithm \cite{pre92}. 
The code has been tested by reproducing the GPE results of 
Ref. \cite{dal96}.

\subsection{Results for hard spheres}

A many-body system of hard spheres is an ideal test-ground for 
various theoretical models, as accurate results for its 
ground-state energy are available from diffusion Monte Carlo 
(DMC) calculations for both homogeneous \cite{gio99} and
inhomogeneous cases \cite{blu01}.

Figure 1 shows the relative differences of the energy functional between GP 
and DMC for $N$ bosons in a spherical harmonic trap characterized by a length 
scale $L_{0}=\sqrt{\hbar/2m\omega}$.
It is clear from these results that the ground state energy of the system
can differ substantially from those predicated by the GP theory,
for either strong confinement (large $a/L_0$) and/or high density (large $N$).
\begin{figure}
\scalebox{0.3}{\includegraphics{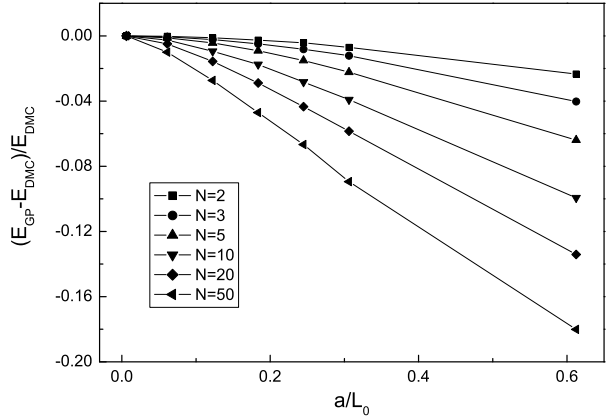}}
\caption{Relative differences between ground state energies predicted by
the GP theory and by diffusion Monte Carlo calculations \cite{blu01}. 
The results are for $N$ bosons in a spherically symmetric trap of size
$L_{0}=\sqrt{\hbar/2m\omega}$. Here $a$ is the scattering length.
\label{Figure1}}
\end{figure}

As shown in Ref.~\cite{blu01}, most of these differences, at least for the range
of parameters considered, can be accounted for by the LHY quantum fluctuation
correction \cite{lhy57}, at a level corresponding to MGPI. 
This is illustrated in Figure~2(a).
\begin{figure}
\scalebox{0.3}{\includegraphics{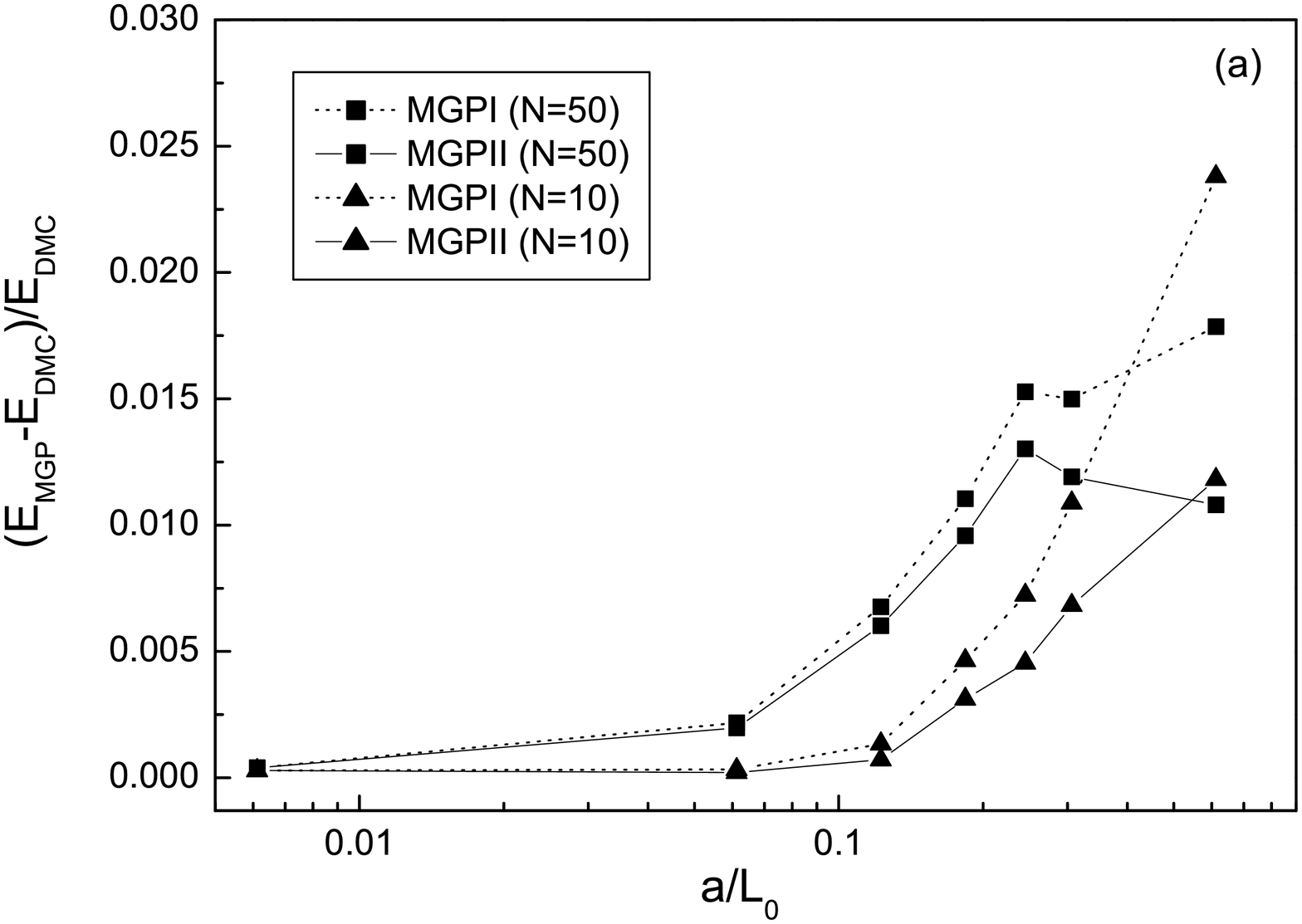}}
\scalebox{0.3}{\includegraphics{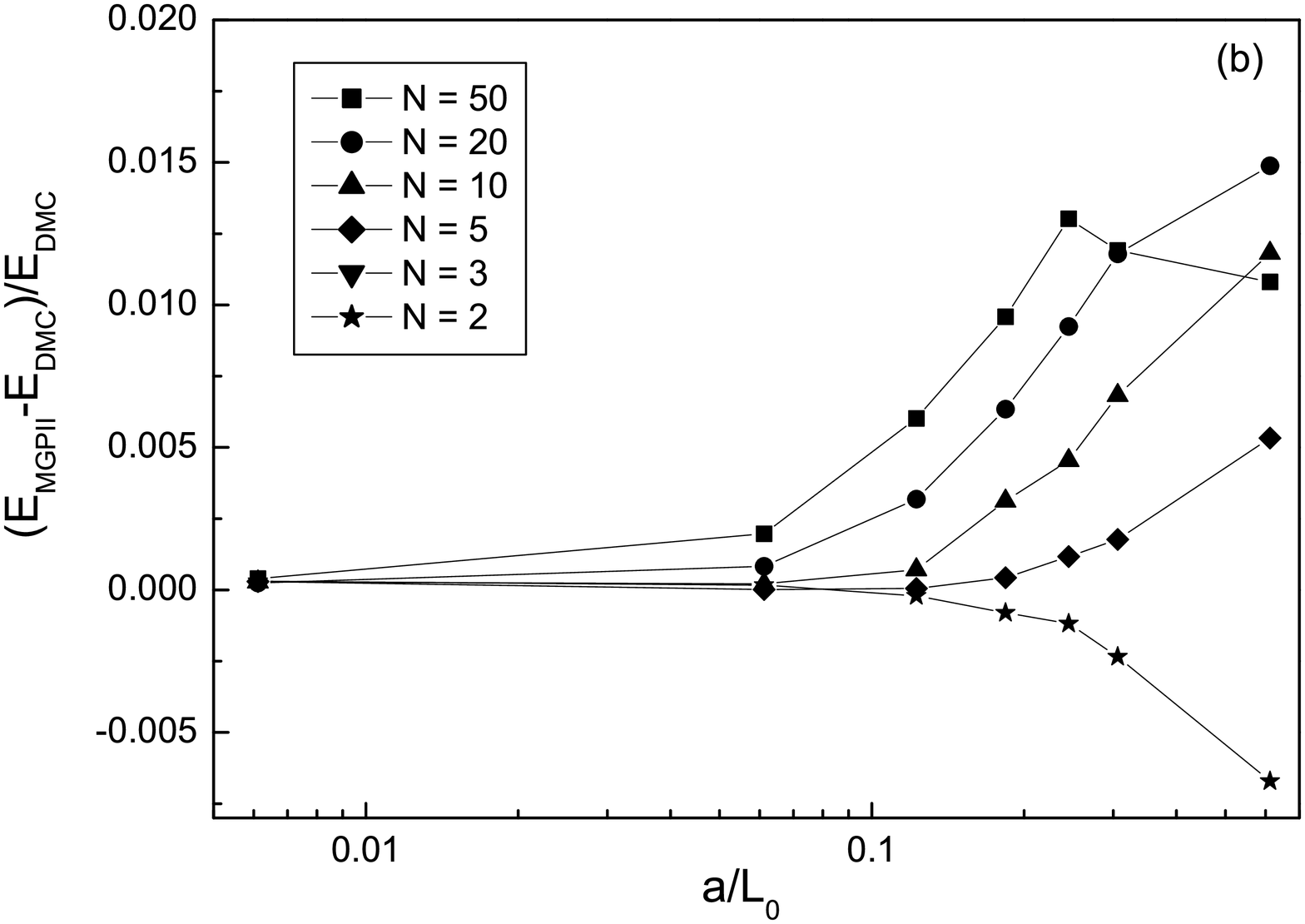}}
\caption{Relative differences between ground state energies predicted by
modified GP theories and by diffusion Monte Carlo calculations \cite{blu01}.
(a) A sample comparison of MGPI and MGPII, both of which lead to much better 
results than the GP theory. (b) More results of MGPII.
Note that the $y$-scale of these figures are about a factor of ten smaller
than that of Figure~1.}
\end{figure}
In the same figure, it is also shown that a consistently better agreement 
with DMC is achieved by MGPII, which includes the shape-dependent confinement correction. 
This means that of the remaining difference between MGPI and DMC,
at least part of it is shape-dependent and requires
a better description of atom-atom interaction. Note that in the case of hard spheres,
$g_2$ is always positive. The sign of the shape-dependent confinement correction 
is then determined by the expectation value of
$\nabla^{2}\left(  \left\vert \phi\right\vert ^{2}\right)$, 
which is negative for the ground state.

More results of MGPII are presented in Figure~2(b). It shows that at the level
of MGPII, the relative difference from DMC has been reduced from up to 20\% for
the GP theory (see Figure~1) to less than 2\% for the range of parameters considered.
 
\subsection{Results for atoms with van der Waals interaction}

With the introduction of $\widehat{H}_\text{mod}$, the ground state energy becomes
dependent upon the shape of the long-range potential through the
relationship between $r_e$ and $a$. For atoms with van der Waals
interaction, $r_e$ and $a$ are related by Eq.~(\ref{eq:reinvr6}).
Figure~3 is a graphic illustration of this relation. Note that $r_e$ diverges
for $a=0$, corresponding to the fact that in this particular case, the energy 
dependence of the scattering amplitude around zero energy cannot be described by
an effective-range expansion \cite{gao00}. 
\begin{figure}
\scalebox{0.3}{\includegraphics{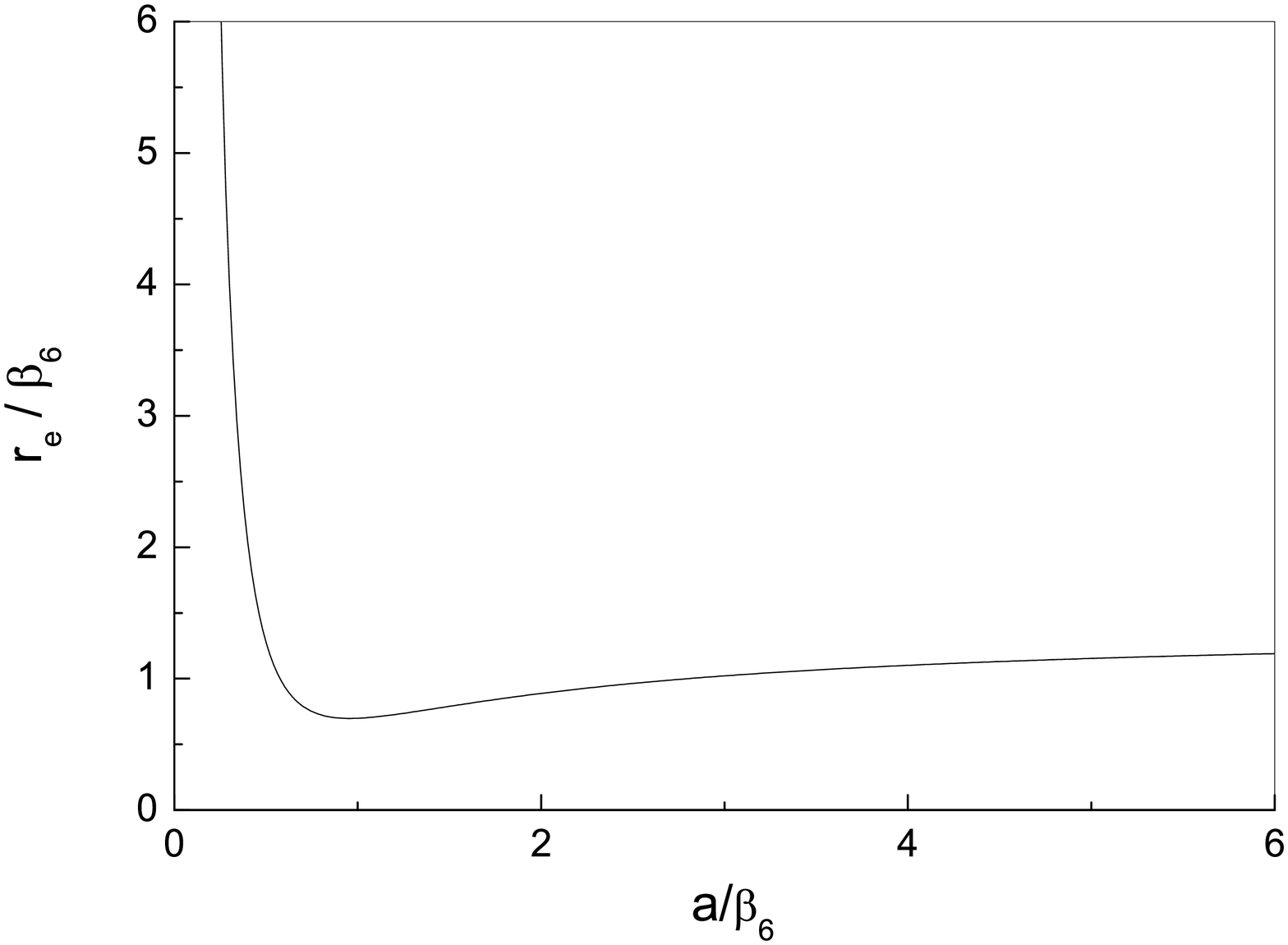}}
\caption{The universal relationship between $r_e/\beta_6$ and $a/\beta_6$ for
atoms with long-range van der Waals interaction \cite{gao98b}.}
\end{figure}
Also note that unlike the case of hard spheres, the $g_2$ for a van der Waals
potential may become negative for sufficiently small
$a/\beta_6$. Consequently, the shape-dependent confinement correction
to the ground state energy may in principle be positive. 
[Both points suggest that something interesting happens around 
$a=0$ that may deserve a separate investigation.]

Figure~4(a) gives a comparison of ground state energies predicted by MGPII and
by the GP theory for a realistic experimental configuration.
Specifically, it is for a $^{85}$Rb condensate in a cylindrical trap with
an aspect ratio of $\varepsilon=6.8/17.5$ \cite{don01}.
A $C_6$ value of 4698 a.u. is used \cite{van02}.
It shows that deviations from the GP theory can become
substantial for either high density and/or strong confinement.
With the tunability of the scattering length via a Feshbach resonance \cite{don01},
such many-body effects beyond the GP theory may soon become observable.
\begin{figure}
\scalebox{0.3}{\includegraphics{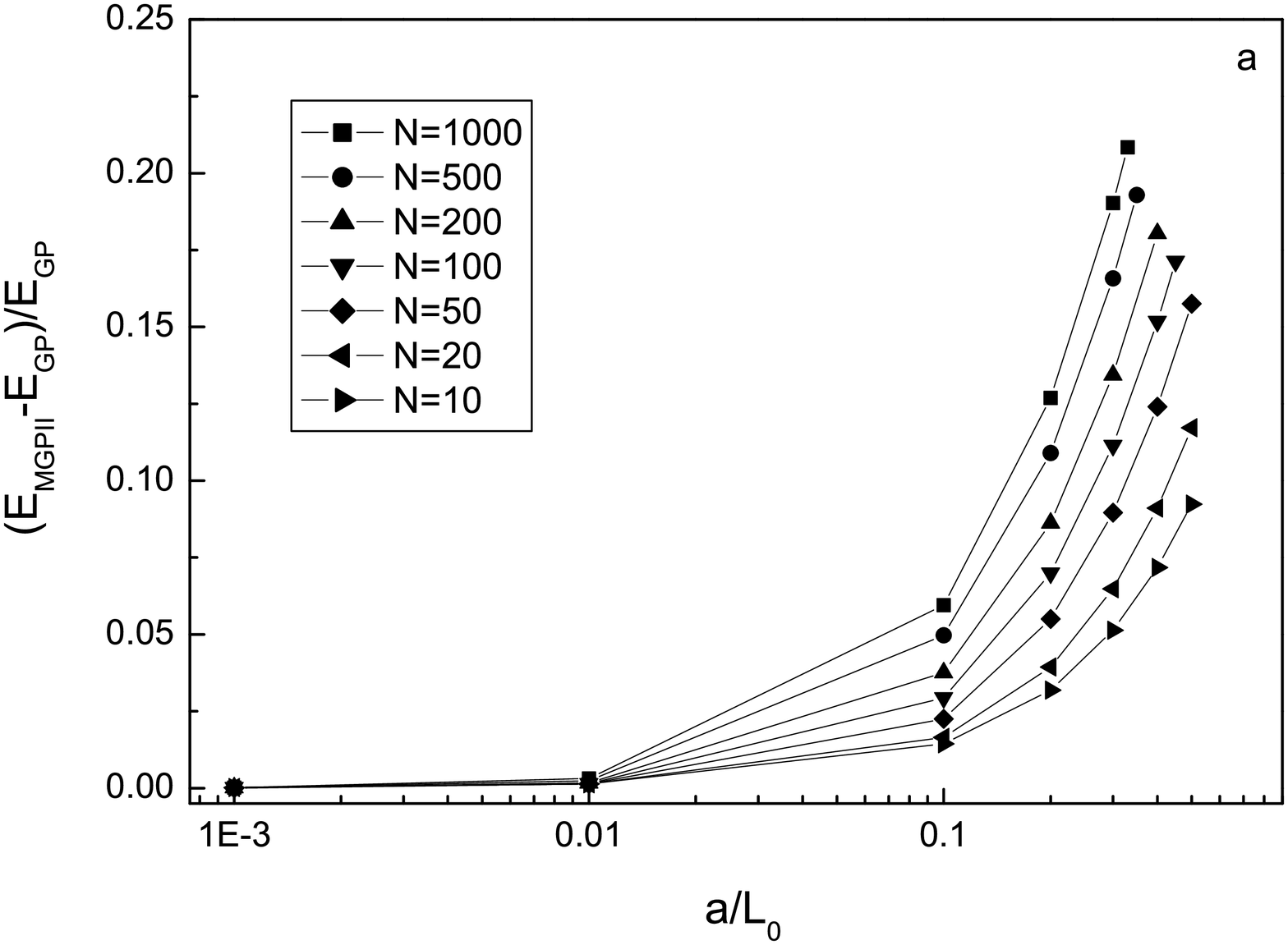}}
\scalebox{0.3}{\includegraphics{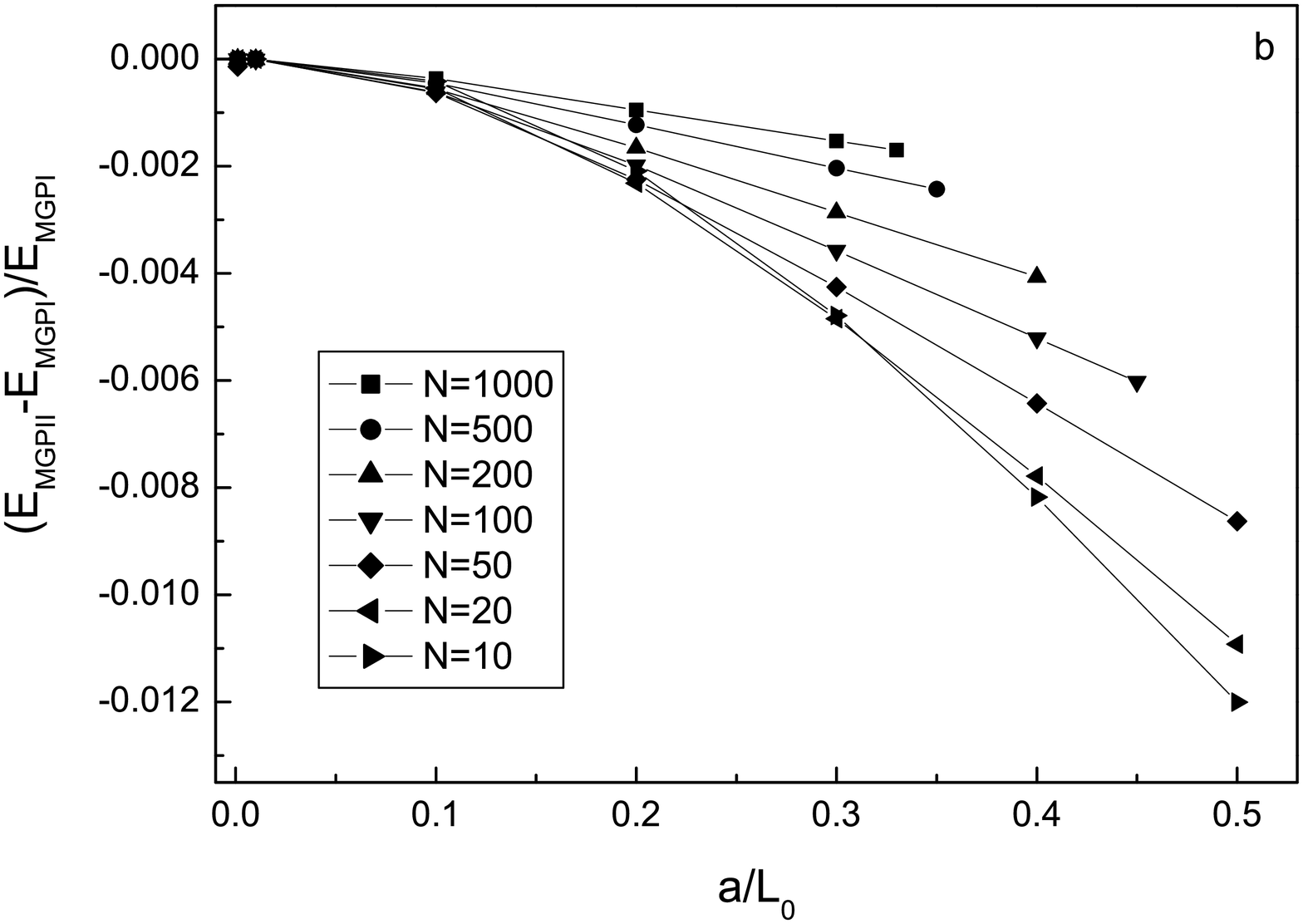}}
\caption{(a) Relative differences between ground state energies predicted by
MGPII and by the GP theory for a $^{85}$Rb condensate in a cylindrical 
trap of aspect ratio $\varepsilon=6.8/17.5$. Here $L_0$ is defined by
the transverse frequency: $L_{0}=\sqrt{\hbar/2m\omega_{\perp}}$.
(b) Relative contribution from the shape-dependent confinement correction.}
\end{figure}
Figure~4(b) shows more explicitly the shape-dependent confinement correction.
Its magnitude increases with $a/L_0$, as expected.
For a fixed $a/L_0$, its relative contribution is more significant for
small $N$ than for large $N$, consistent with our earlier discussion.
The corrections are mostly negative (except for very small $a$) since
$g_2$ is greater than zero for most of the data points shown.
[Because $L_0=34850$ a.u. is much greater than $\beta_6=164.2$ a.u., 
most of the data points correspond to $a/\beta_6\gg 1$. 
The $r_e$ is then approximately a constant  
$r_e\approx (3\pi)^{-1}[\Gamma(1/4)]^2\beta_6$, 
and the corresponding $g_2$'s are positive.]

\section{Conclusion}

In conclusion, an effective interaction and the corresponding modified GP
equation have been proposed that take into account the energy dependence
of the two-body scattering amplitude through an effective-range expansion.
The resulting theory, called MGPII, leads to better agreements with  
diffusion Monte Carlo calculations \cite{blu01} than either the GP theory, or
MGPI that considers only the quantum fluctuation correction.
The theory expands, considerably, the parameter space 
(specified by $na^3$ and $a/L_0$),
in which a GP type of formulation can be applied.
It introduces the concept of shape-dependent confinement correction and 
shows that for a fixed confinement, its relative contribution
is more significant for small $N$ than for large $N$.

Finally, we point out that both MGPI and MGPII are
perturbative expansions around the GP theory. They will
eventually fail for sufficiently large $na^3$ and/or sufficiently
strong confinement. For high density, shape dependence of a different 
origin will eventually emerge from two-body correlations at short 
length scales \cite{cow02,gao03}. 
For strong confinement, the local density approximation, implicit in both
MGPI and MGPII, will eventually fail. Exploration into those regimes will require
theories that differ much more substantially from the GP formulation.

\begin{acknowledgments}
The authors are deeply indebted to Dr. Doerte Blume for providing her results 
of DMC calculations. This work was
supported by the National Natural Science Foundation of China No. 19834060 and
the Key Project of Knowledge Innovation Program of Chinese Academy of Science
No. KJCX2-W7. BG is also supported in part by the US National Science Foundation.
\end{acknowledgments}

\end{document}